\documentclass[12pt,preprint]{aastex}

\newcommand{\Mjup}{$M_\mathrm{J}$}
\newcommand{\Rjup}{$R_\mathrm{J}$}

\newcommand{\spitzer}{\emph{Spitzer}}
\newcommand{\microns}{$\mu$m}
\newcommand{\etal}{et al.}

\newcommand{\ratiothree}{(R_p/R_\star)_{5.8\,\mu\mathrm{m}}}
\newcommand{\ratioone}{(R_p/R_\star)_{3.6\,\mu\mathrm{m}}}

\begin{document}

\title{A \emph{Spitzer} Search for Water \\ in the Transiting Exoplanet HD\,189733\lowercase{b} \\
{\scriptsize ---Accepted for publication in \emph{ApJ Letters}---}}
\author{David Ehrenreich, Guillaume H\'ebrard, Alain~Lecavelier des Etangs, David~K.~Sing,\\ Jean-Michel~D\'esert,
Fran\c{c}ois~Bouchy, Roger~Ferlet, and Alfred~Vidal-Madjar}

\affil{Institut d'astrophysique de Paris, CNRS (UMR 7095), Universit\'e Pierre
et Marie Curie, 98 bis boulevard Arago, 75014 Paris, France}

\begin{abstract}
We present \spitzer\ \emph{Space Telescope} observations of the extrasolar
planet HD\,189733b primary transit, obtained simultaneously at 3.6 and
5.8~\microns\ with the Infrared Array Camera. The system parameters, including
planetary radius, stellar radius, and impact parameter are derived from fits to
the transit light curves at both wavelengths. We measure two consistent
planet-to-star radius ratios, $\ratioone = 0.1560 \pm 0.0008\textrm{(stat)} \pm
0.0002\textrm{(syst)}$ and $\ratiothree = 0.1541 \pm 0.0009\textrm{(stat)} \pm
0.0009\textrm{(syst)}$, which include both the random and systematic errors in
the transit baseline. Although planet radii are determined at 1\%-accuracy, if
all uncertainties are taken into account the resulting error bars are still too
large to allow for the detection of atmospheric constituants like water vapour.
This illustrates the need to observe multiple transits with the longest
possible out-of-transit baseline, in order to achieve the precision required by
transmission spectroscopy of giant extrasolar planets.
\end{abstract}


\section{INTRODUCTION}

\label{sec:intro} During a planetary transit, the eclipsed light from the star
filters through the atmospheric limb of the planet. Transmission spectroscopy
of this light lead to detect and probe the deep and upper-escaping atmospheres
of HD\,209458b (Charbonneau \etal\ 2002; Vidal-Madjar \etal\ 2003, 2004;
Ballester \etal\ 2007). Richardson \etal\ (2006) obtained the first infrared
(IR) transit measurement for this planet and found its radius at 24~\microns\
consistent with the visible radius. Based on planetary radius measurements by
Knutson \etal\ (2007a) at optical wavelengths, Barman (2007) claimed the
identification of water in the planet atmosphere.

HD\,189733b, discovered by Bouchy \etal\ (2005), is orbiting a small, close,
and bright main sequence K star, thus giving the deepest transit occultation
ever detected ($\sim$2.5\%). The planet has a mass $M_p = 1.13$~Jovian mass
(\Mjup) and its radius in the visible is $R_p = 1.16$~Jovian radius (\Rjup;
Bakos \etal\ 2006; Winn \etal\ 2007). Fortney \& Marley (2007) suggested a
possible water detection in this planet, yielding from \emph{Spitzer}
observations of an anti-transit, whereas Knutson \etal\ (2007b) obtained the
planet-to-star radius ratio at 8~\microns\ and found
$(R_p/R_\star)_\mathrm{8\,\mu m} = 0.1545 \pm 0.0002$.

Here we describe the \spitzer\ observations collected during the primary
transit of HD\,189733b in order to measure its radius at two different IR
wavelengths and search for atmospheric water (H$_2$O). Models of the IR
transmission spectrum of this planet (Tinetti \etal\ 2007a,b) have shown that
\spitzer\ is well suited to probe the planet atmospheric composition, in
particular by comparing two photometric bands, centered at 3.6 and
5.8~\microns. The absorption by H$_2$O should give a difference in the spectral
ratios measured at those two wavelengths of $\Delta_{\Re'}({\rm H_2O}) \equiv
(\Re'_{5.8\mathrm{\mu m}} - \Re'_{3.6\mathrm{\mu m}}) / \Re'_{3.6\mathrm{\mu
m}} \sim 1.7$--$3.4\%$, depending on the set of H$_2$O absorption cross-section
coefficients used for the calculation, and where $\Re' \approx (R_p /
R_\star)^2$, as defined by Brown (2001). This corresponds to a predicted
planetary radius relative difference due to absorption by H$_2$O of
$\Delta_{R}({\rm H_2O}) \sim 0.85$--$1.7\%$.

\section{OBSERVATIONS}
\label{sec:observations}

We observed HD\,189733 on 2006 October 31, during a primary transit of its
planet with the Infrared Array Camera (IRAC, Fazio \etal\ 2004). Our 4.5-h
observations covers the 1.8-h transit of HD\,189733b. We used only one IRAC
channel pair to avoid repointing the telescope during the observations: the
0.75-\microns-wide channel~1 centered on 3.6~\microns, and the
1.42-\microns-wide channel~3 centered on 5.8~\microns. We did not dither the
pointing in order to keep the source on a particular position of the detector
and increase the photometric accuracy.

The observations were split in 1936 consecutive sub-exposures, each integrated
over 0.4 and 2~s (frame times) for channels~1 and~3, respectively. The short
exposure times in IRAC `stellar mode' avoid the saturation of the detector due
to HD\,189733, a $K=5.5$ magnitude star. We used the flat-fielded,
cosmic-ray-corrected, and flux-calibrated data files provided by the \spitzer\
pipeline.

\section{DATA ANALYSIS}
\label{sec:data_analysis}

\subsection{Photometry and background}

\label{sec:method} To obtain a transit light curve from the two-dimensional
1936 images, we calculate a weighted aperture photometry by summing the
weighted background-subtracted flux on each pixel within an aperture of given
radius $r$ (Horne 1986; Naylor 1998). The optimal weighting on a pixel is $P /
\sigma^2$, where $P$ and $\sigma$ are the values of the point spread function
(PSF) and photon noise for this pixel. The PSF is estimated in each channel and
for each pixel as the median of the background-subtracted fluxes. Finally, the
estimated error on the weighted integrated flux is calculated as the
square-root of the weighted photon-noise quadratic sum; it remains a constant
throughout the time series.

To estimate the sky and instrumental background for each exposure, we
calculated the mean value of the image in an annulus centered on the star with
inner and outer radii of 16 and 18 pixels, respectively. Different ring sizes
were tested to check that (1) the stellar PSF does not contaminate the
background and (2) other field stars contribution is minimized. Typical
background estimates are $\sim 0.05$ and $0.1$--$0.2$~mJy per pixel in
channels~1 and~3, respectively. They are $\sim 10^4$ times less than stellar
flux integrated over the 113-pixel photometric aperture.

The initial weighted flux time series were extracted with an aperture radius $r
= 6$~pixels. The raw weighted light curves in channels~1 and~3 are plotted in
Fig.~\ref{fig:time_series}. Beyond the main trend due to the expected planetary
transit, with an occultation depth of more than 2\% during about 6,500~s,
additional effects pollute the signal. In both channels, there is a strong
decrease of the flux during the first $\sim 1,000$~s of observations. Most
noticeable in channel~1 at 3.6~\microns, the star is close to nominal
saturation limits and the light curve presents large fluctuations, attributed
to the `pixel-phase effect'. A close look to the 2D images obtained in
channel~3 at 5.8~\microns\ reveals a contamination of the photometry caused by
the `bandwidth effect'. The light curve baseline is also affected in this
channel by a decreasing drift. These effects and their corrections are now
further detailed.

\subsection{Instrumental artifacts}
\label{sec:effects}

\paragraph{Saturation.}
\label{sec:saturation} The flux of HD\,189733 is 1,700~mJy at 3.6~\microns.
This is about twice the maximum recommended (and conservative) point source
value for a frame time of 0.4~s in stellar mode.\footnote{See
http://ssc.spitzer.caltech.edu/irac/sat.html.} Looking into the raw data from
channel~1, we found that the brightest pixel of the stellar PSF is above the
detector array saturating value ($DN = 30,000$) \emph{only} during the
$\sim100$ first sub-expsosures. Those are already discarded for the data
reduction. In the following sub-exposures, the flux remains below the
saturation limit, in the linear regime.

\begin{figure}
\resizebox{\hsize}{!}{\includegraphics{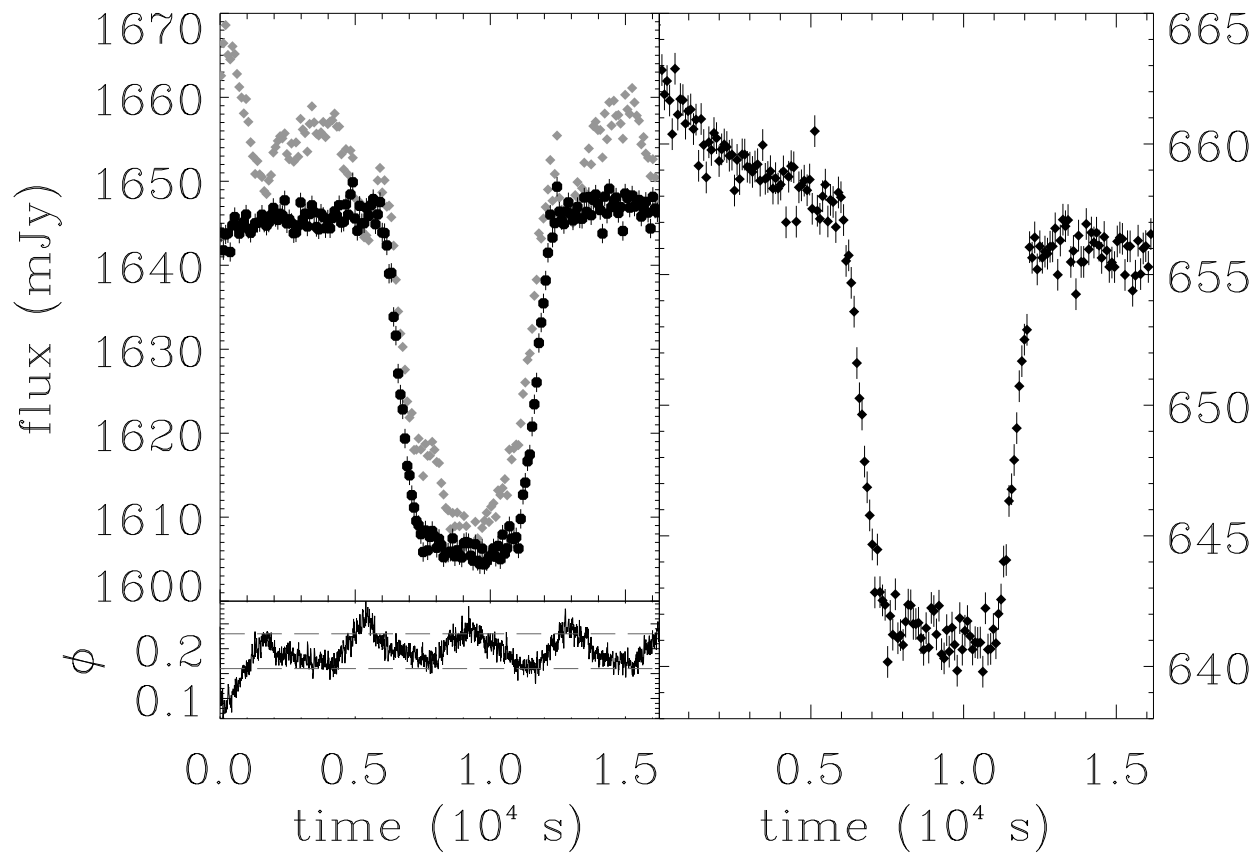}} \caption{Weighted light curves
in channel~1 at 3.6~\microns\ (left) and channel~3 at 5.8~\microns\ (right).
Data are rebinned by 10. The raw light curve at 3.6~\microns\ (grey diamonds)
has to be corrected for large fluctuations correlated to the `pixel phase',
plotted in the left lower panel. Those exposures with extreme pixel phases
(beyond the dashed lines) are rejected. The corrected light curve is
overplotted as black circles in the upper panel.} \label{fig:time_series}
\end{figure}

\paragraph{Pixel-phase effect at 3.6~\microns.}
\label{sec:pixel-phase} The telescope jitter and intra-pixel sensitivity
variations for the observation of a bright star are likely responsible for the
large fluctuations seen in the light curve obtained at 3.6~\microns\ in
channel~1. These fluctuations are correlated to the pixel-phase variation,
whose effect is most severe in channel~1. A description of this effect and a
correction method are given in the IRAC data handbook (Reach et al.\ 2006,
p.~50). The method, also reported by Morales-Calder\'on \etal\ (2006), consists
in calculating a pixel-phase dependent correction on the flux, $F_w^{\rm cor} =
F_w \left[ 1 + k \left( 1 / 2\pi - \phi \right) \right]^{-1}$, where the pixel
phase is $\phi = [(x-x_0)^2 + (y-y_0)^2]^{1/2}$, $(x,y)$ is the centroid of the
point source, and $x_0$ and $y_0$ are the integer pixel numbers containing the
source centroid. The optimized correction is determined by iteratively fitting
the out-of-transit flux baseline. The pixel phase variations and the raw and
corrected light curves are plotted in Fig.~\ref{fig:time_series} (left). The
relative difference introduced by this correction on the value of $\ratioone$
can be estimated to $\Delta_{R}^\mathrm{3.6\mu m}(\mathrm{phase}) \sim 2$--3\%.

\paragraph{Bandwidth effect at 5.8~\microns.}
\label{sec:bandwidth} The bandwidth effect reportedly affects those IRAC
channels fitted with detector arrays made of arsenic doped silicon (Si:As),
such as channel~3. The IRAC data handbook (Reach et al.\ 2006, p.~24) describes
it as decaying echoes 4, 8, and 12 columns to the right of a bright or
saturated pixel. HD\,189733 is no brighter than 700~mJy at 5.8~\microns,
whereas the maximum unsaturated point source brightness at this wavelength and
for 2-s frame time is 1,400~mJy.$^1~$ Yet, the pixel located 4 columns to the
right of the stellar maximum is anomalously bright in all exposures and appears
as a peak in the wing of the stellar PSF, $\sim 2$--3 times brighter than
closer-to-the-centroid adjacent pixels. Therefore, we rejected this suspicious
pixel from the aperture photometry integration, as recommended by IRAC status
reports.\footnote{See the October and December 2005 IRAC status reports at
https://lists.ipac.caltech.edu/mailman/htdig/irac-ig.} This decreases the value
obtained for $\ratiothree$, and the relative difference between the corrected
and uncorrected values is $\Delta_{R}^\mathrm{5.8\mu m}({\rm band}) \sim 1\%$.
The bandwidth effect could typically lead to obtain a planetary radius
systematically larger at 5.8 than at 3.6~\microns\ and mimic an absorption due
to atmospheric water, hence leading to a false detection.

\begin{figure}
\resizebox{\hsize}{!}{\includegraphics{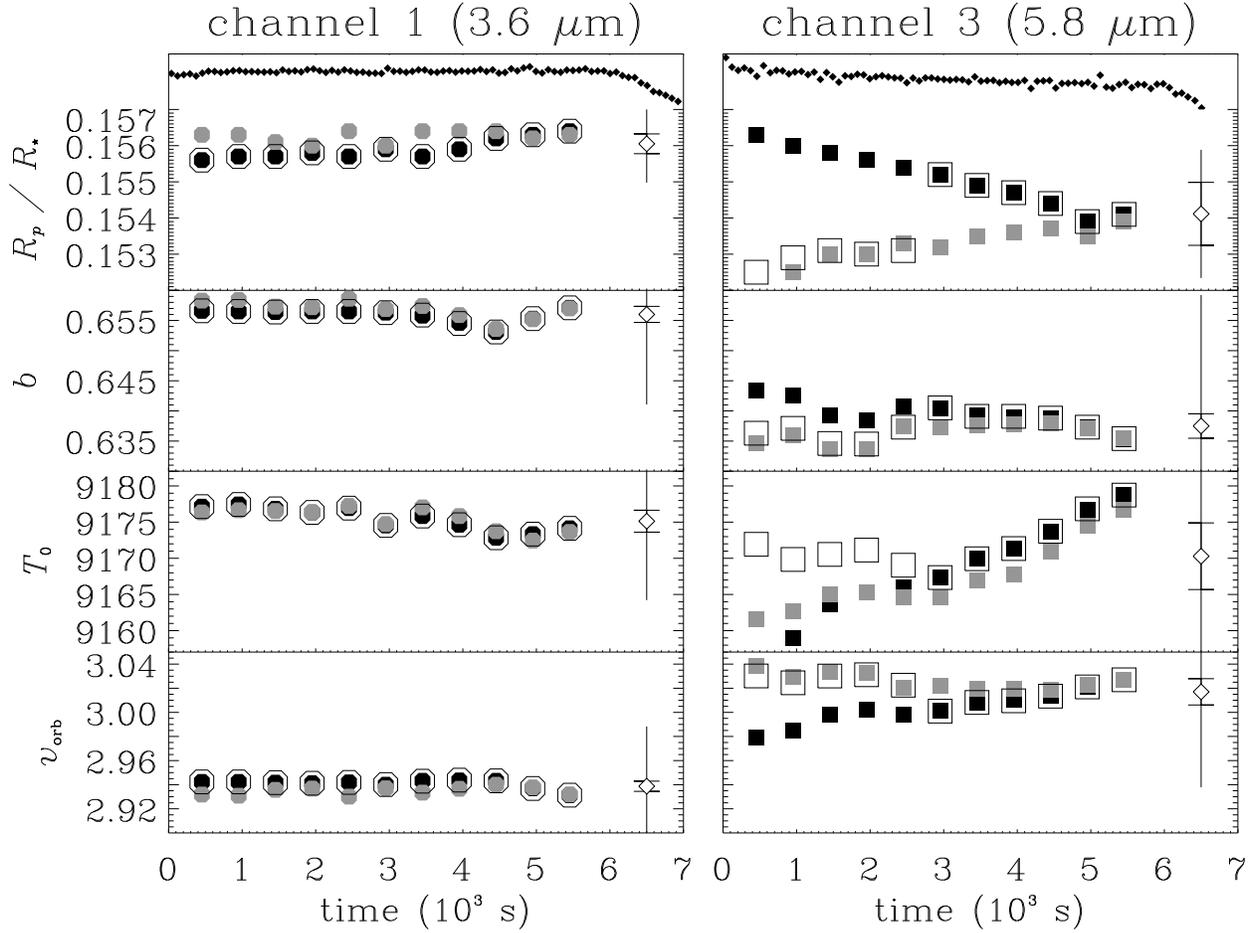}} \caption{System parameters
obtained in channels~1 (left) and~3 (right), as a function of the time from
which the transit light curve is fitted. The light curves in both channels are
plotted in the upper panels until the ingress. The parameters shown are, from
top to bottom, $R_p/R_\star$, $b$, $T_0$, and $v_\mathrm{orb}$. To correct for
the decreasing drift in channel~3, the transit light curve model can include a
linear (black), quadratic (grey), or exponential (empty symbols) out-of-transit
baseline. For consistency, we also applied these fitting tests to channel~1.
The dispersion observed in the results is accounted for by choosing the mean of
each sample (empty diamonds) and adding a systematic uncertainty equal to the
standard deviation in each sample. The error bar represented in each panel
accounts for the statistical \emph{and} systematic errors. The contribution of
the systematics is indicated by the horizontal bars.} \label{fig:time_start}
\end{figure}

\paragraph{Drift of the flux at 5.8~\microns.}
\label{sec:drift} A non-linear decreasing drift can be seen in the channel-3
light curve (Fig.~\ref{fig:time_series}, right). After steeply decreasing, the
drift seems to set as a nearly linear trend after 2,500~s. This gives us the
choice to drop the exposures before that time and linearly fit the
out-of-transit baseline after, or keep most of the exposures at the beginning
of the observations and perform either a quadratic or exponential fit to the
baseline.
We tested the influence of both the polynomial fitted to the baseline and the
number of exposures dropped from the beginning of the observations on the
system parameters yielding from the fitting procedure. To this purpose, the
time $t_s$ defining the start of the fit was set as a free parameter. For
consistency, the same tests were performed in channel~1, and their results are
plotted in Fig.~\ref{fig:time_start}. A large dispersion of values is obtained,
especially for $\ratiothree$ in channel~3. Depending on the fit parameters, the
dispersion obtained are $\Delta_{R}^\mathrm{3.6\mu m}({\rm drift}) \sim 0.1\%$
and $\Delta_{R}^\mathrm{5.8\mu m}({\rm drift}) \sim 0.6\%$ at 3.6 and
5.8~\microns, respectively.
The limited knowledge of the baseline exact level during the transit introduces
systematic uncertainties in the determination of the system parameters. These
uncertainties are further taken into account as systematic errors.

\subsection{Determination of the system parameters}
\label{sec:system_parameters}

\paragraph{Selection of sub-exposures.}
\label{sec:selectexpo} We made a selection within the 1936~sub-exposures to
obtain the best possible photometry. Sub-exposures where the aperture contains
at least one pixel flagged by the \spitzer\ pipeline, are removed from our time
series. We did not apply such selection to one particular pixel always present
in the wing of the PSF in channel~1: it is systematically flagged as having a
`photometric accuracy unacceptably low,' which is verified when compared to
adjacent pixels. However, we found no significant differences when including or
excluding it from the aperture photometry. We also removed the dozen
sub-exposures in each channel where the integrated photometry of HD\,189733
presents strong and isolated variations. Finally, we kept in channel~1 the
exposures where the pixel phase was between $0.16 < \phi < 0.23$ (see
Fig.~\ref{fig:time_series}), and rejected the others in order to minimize the
influence of residuals from the correction for the pixel-phase effect. As a
result, when cutting out the first 500~s of data after the beginning of the
observations, we consider 75 and 96\% of the total number of exposures in
channels~1 and 3, respectively.

\paragraph{Fitting the transit light curves.}
\label{sec:fit_method} The transit light curves at 3.6 and 5.8~\microns\ are
fitted with a procedure based on the analytical model of Mandel \& Agol (2002),
which includes the effect of limb-darkening. The procedure is able to fit
either linear, quadratic, or exponential baselines. The resulting parameters of
the fit at each wavelength are the planet-to-star radius ratio $R_p/R_\star$,
the impact parameter $b$ in units of stellar radii, the orbital velocity
$v_\mathrm{orb}$ in units of stellar radii which, because the planet orbital
period is known to high accuracy ($2.218574$~days, according to H\'ebrard \&
Lecavelier des Etangs 2006), can be converted into $R_\star M_\star^{-1/3}$,
where $M_\star$ is the stellar mass, and the heliocentric transit central time
$T_0$. The best fits obtained are plotted in Fig.~\ref{fig:fits}.

\paragraph{Limb darkening effect.}
\label{sec:limb-darkening} The contribution of limb-darkening to the transit
light curve is calculated using a non-linear limb-darkening law (Mandel \&
Algol 2002) which has four wavelength-dependent coefficients. These
coefficients were fitted using a Kurucz (2005) stellar model
($T_\mathrm{eff}=5,000$~K, $\log{g}=4.5$, solar abundance), which closely
matched the observed parameters of HD\,189733, at 17 different angles from
center to limb. The stellar model was convolved, at each angle, with the IRAC
photometric bandpasses before fitting the non-linear law. We found the
coefficients $C_1$, $C_2$, $C_3$, and $C_4$ of the law to be $0.6023$,
$-0.5110$, $0.4655$, and $-0.1752$ at 3.6~\microns, and $0.7137$, $-1.0720$,
$1.0515$, and $-0.3825$ at 5.8~\microns. The uncertainty in the limb-darkening
coefficients has no impact on the results.
However, the uncertainty in the impact parameter introduces an uncertainty in
the limb darkening amplitude and, therefore, an uncertainty in the measured
planetary radius. The relative radius difference at 3.6~\microns\ due to the
limb-darkening effect is $\Delta_{R}^\mathrm{3.6\mu m}({\rm limb}) \sim 1\%$
and $\Delta_{R}^\mathrm{5.8\mu m}({\rm limb}) \sim 0.3\%$ at 3.6 and
5.8~\microns, respectively. The limb-darkening effect can be appreciated in the
bottom panel of Fig.~\ref{fig:fits}.
\paragraph{Statistical error bars.}
The statistical error bars on the parameters are calculated with the $\Delta
\chi^2$ method described by H\'ebrard \etal\ (2002). The quality of the fit is
given by the value of $\chi^2/n$, where $n$ is the degree of freedom of the
light curve. Assuming we are limited by the photon noise, we find $\chi^2/n$ of
$\sim 1.5$ and $1.3$ at 3.6 and 5.8~\microns, respectively. We thus scaled the
uncertainties larger by factors of $\sqrt{1.5} = 1.22$ and $\sqrt{1.3} = 1.14$
to obtain $\chi^2/n \sim 1$ in both channels.
Using various models and starting time for the baseline gives similar $\chi^2$
values, showing that the light curve does not contain enough information to
constrain that source of uncertainty.

\paragraph{Systematics.}
The effects described above all introduce systematics that are clearly not
negligible compared to the predicted radius differences due to atmospheric
water, $\Delta_{R}({\rm H_2O}) \sim 0.85$--$1.7\%$. Limb-darkening effect,
introducing $\Delta_{R}^\mathrm{3.6\mu m}({\rm limb}) \sim 1\%$ in channel~1
and $\Delta_{R}^\mathrm{5.8\mu m}({\rm limb}) \sim 0.3\%$ in channel~3 is dealt
with as described in the previous section. Tests shown that we are able to
fairly correct for $\Delta_{R}^\mathrm{3.6\mu m}({\rm phase}) \sim 2\%$ and
$\Delta_{R}^\mathrm{5.8\mu m}({\rm band}) \sim 1\%$. On the other hand,
additional uncertainty have to be introduced to properly handle the systematics
linked to the drift in the flux seen in channel~3 and, to a lesser extent, in
channel~1. Indeed, basing ourselves on the similar reduced $\chi^2$ obtained
when fitting the baseline with different polynomials, we cannot choose one of
the sets of system parameters rather than another. Besides, in the absence of a
`plateau' in the plot of $\ratiothree$ vs.\ $t_s$, we cannot either favor one
solution based on the time $t_s$ chosen to start the fitting procedure.
After removing the solutions corresponding to the extreme values of $t_s$, we
thus set the value of each parameter, in each channel, to the mean of each
sample of solutions. The uncertainties on the obtained values should reflect
the dispersion observed. Therefore, we set a systematic error bar on each
parameter, equal to the standard deviation in each sample of solutions.

\section{RESULTS AND DISCUSSION}
\label{sec:results}

\begin{figure}
\resizebox{\hsize}{!}{\includegraphics{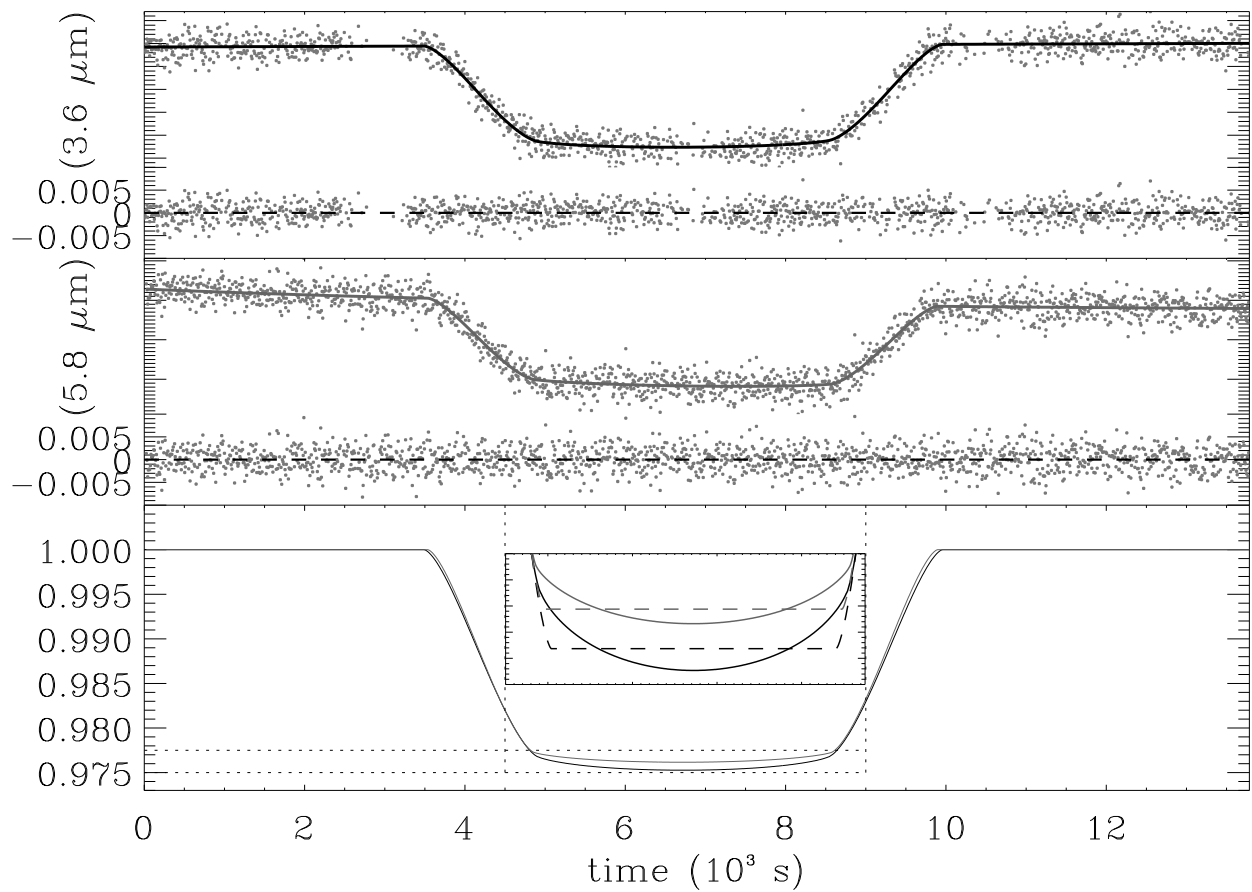}} \caption{Final light curves of
HD\,189733 during the Oct.~31st 2006 transit at 3.6 (top) and 5.8~\microns\
(middle). Fits to the light curves (thick lines) are calculated from the system
parameters given in Table~\ref{tab:param}. The residuals are shown below each
fitted light curve. The lower panel shows a comparison between the two fits.
The inlet contains a zoom on the transit bottom, where our best-fits obtained
without limb-darkening are superimposed (dashed lines).} \label{fig:fits}
\end{figure}

The quality of the method is confirmed by the good agreement between system
parameters independently obtained with the best fits to the light curves at
both wavelengths (see Fig.~\ref{fig:fits}; the values are reported in
Table~\ref{tab:param}). We measure consistent planet-to-star radius ratios of
$0.1560 \pm 0.0008 \textrm{(stat)} \pm 0.0002 \textrm{(syst)}$ and $0.1541 \pm
0.0009 \textrm{(stat)} \pm 0.0009 \textrm{(syst)}$, at 3.6 and 5.8~\microns,
respectively. Using the notation introduced above, we find $\Delta_{R}({\rm
obs}) = -0.84 \pm 1.00 \textrm{(stat)} \pm 0.84 \textrm{(syst)}$\%.

Tinetti \etal\ (2007a,b) estimated that the presence of H$_2$O in the
atmosphere of the planet would result in a radius at 5.8~\microns\ being
$\Delta_{R}({\rm H_2O}) \sim 0.85$--$1.7\%$-larger than at 3.6~\microns. Our
result is 0.9$\sigma$ away from the lower bound of the predictions interval; it
is 1.4$\sigma$ away from the upper bound. The present results are also
significantly different from Tinetti \etal 's (2007b), which are obtained from
a preliminary analysis of the same data set. The difference is mainly due to
several effects taken into account and discussed in the present work: the
bandwidth effect, the determination of the light curve baseline, and the limb
darkening. All these effects have the same order of magnitude as the predicted
H$_2$O absorption and could cause a false positive detection. In particular,
the limb darkening -- in this system with a large $b$ -- makes the occultation
depth $\Re' \neq (R_p/R_\star)^2$ and impacts on the error budget. Since this
effect also depends on the wavelength, it is inaccurate to base the detection
of an atmospheric signature only on the raw difference of occultation depths.

\begin{figure}
\resizebox{\hsize}{!}{\includegraphics{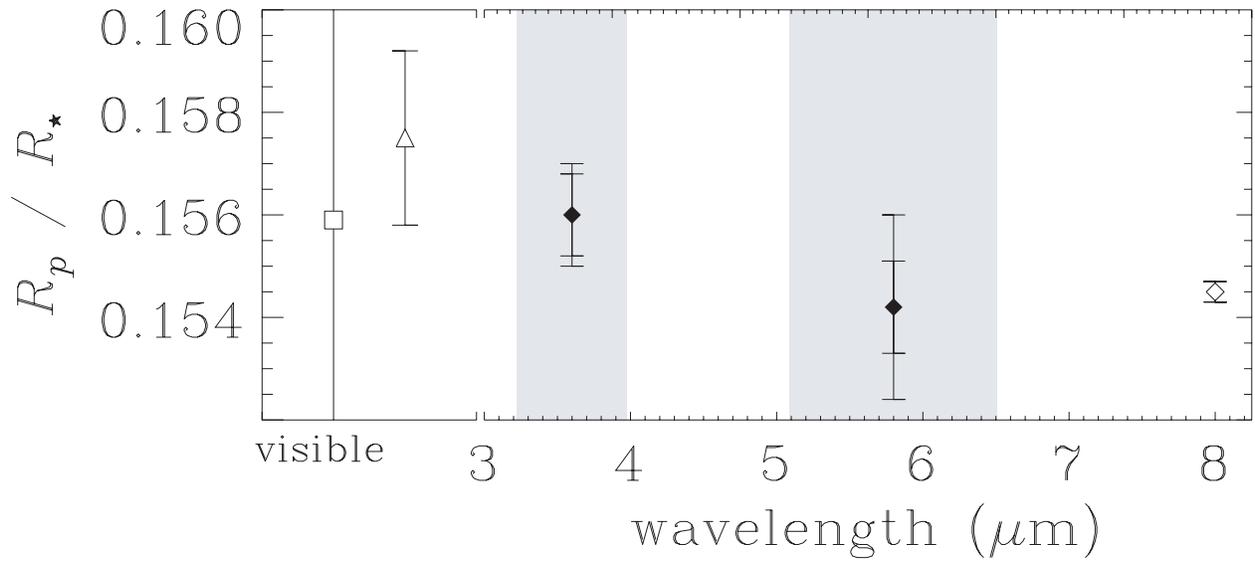}} \caption{Radius of the
planet as a function of wavelength (upper panel), expressed in stellar radii.
The two measurements at 3.6 and 5.8~\microns\ are represented (filled diamonds)
in the near IR. Both IRAC bandpasses are also indicated (grey areas). Previous
measurements in the visible (Bakos \etal\ 2006 [square]; Winn \etal\ 2007
[triangle]) and in the IR (Knutson \etal\ (2007b [empty diamond]) are shown for
comparison.} \label{fig:radii}
\end{figure}

Most recent radius measurements for HD\,189733b are plotted in
Fig.~\ref{fig:radii}. A particular comparison between the system parameters we
derived in the IR and those derived by Winn \etal\ (2007) shows that $R_p$, $b$
and $R_\star$ values at 3.6 and 5.8~\microns\ are consistent with the visible
values (see Table~\ref{tab:param}). Our two radius measurements are also
compatible with the value derived at 8~\microns\ by Knutson \etal\ (2007b),
which has a rather small uncertainty compared to ours. Our statistical
uncertainty is of the same order than the one derived by Winn \etal\ (2007).
During the last stage of the publication of this work, new measurements in the
visible have been reported by Pont \etal\ (2007) using the \emph{Hubble Space
Telescope}. Their derived system parameters are within 1 to 2$\sigma$ from
ours; this marginal disagreement might be explained by stellar spots in such an
active K-type star observed at different epochs.

More generally, the consistency between visible and IR radii for other
extrasolar planets, like HD\,209458b -- measured in the visible by Knutson
\etal\ (2006a) and at 24~\micron\ by Richardson \etal\ (2006) -- or GJ~436b
 -- measured in the visible by Gillon \etal\ (2007a) and at 8~\micron\ by Gillon
\etal\ (2007b) and Deming \etal\ (2007) --, shows that we do not yet achieve
radius determination with enough accuracy in the IR to allow for a
spectroscopic characterization of close-in atmospheres. The accuracy required
($\sim 10^{-4}$) could be obtained by observing several transits with the
longest possible out-of-transit baseline, in order to better constrain the
systematics in the transit curve. New \spitzer/IRAC observations of HD\,189733b
at 3.6, 4.5, and 8~\microns\ should allow the present results to be better
constrained.

\acknowledgements We thank the anonymous referee who greatly contributed to
improve the paper, as well as S.~Carey and V.~Meadows for their help. D.K.S.\
is supported by CNES. This work is based on observations made with the
\emph{Spitzer Space Telescope}, which is operated by the Jet Propulsion
Laboratory, California Institute of Technology under a contract with NASA.

\clearpage

\begin{deluxetable}{lrrr}
    \tabletypesize{\scriptsize}
    \tablecaption{System Parameters\label{tab:param}}
    \tablehead{
    \colhead{Parameter}                                   & \colhead{Visible\tablenotemark{b}} & \colhead{3.6~\microns\tablenotemark{c}}                       & \colhead{5.8~\microns\tablenotemark{c}}}
    \startdata
    $R_p / R_\star$                                       & $0.1575 \pm 0.0017$                & $0.1560 \pm 0.0008 \pm 0.0002$ (0.8$\sigma$)\tablenotemark{d} & $0.1541 \pm 0.0009 \pm 0.0009$ (1.4$\sigma$)\tablenotemark{d} \\
    $b$                                                   & $0.658  \pm 0.027$                 & $0.656  \pm 0.014  \pm 0.001 $ (0.1$\sigma$)\tablenotemark{d} & $0.638  \pm 0.020  \pm 0.002 $ (0.6$\sigma$)\tablenotemark{d} \\
    $(R_\star / R_\odot)(M_\star/0.82M_\odot)^{-1/3}$     & $0.753 \pm 0.025$                  & $0.747  \pm 0.011  \pm 0.001 $ (0.2$\sigma$)\tablenotemark{d} & $0.728  \pm 0.016  \pm 0.003 $ (0.8$\sigma$)\tablenotemark{d} \\
    $T_0$\tablenotemark{a} (s)                            &                                    & $53,214 \pm 9      \pm 2     $                                & $53,218 \pm 11     \pm 5     $                                \\
    \enddata
    \tablenotetext{a}{Given as $T_\mathrm{UTCS} - 215,500,000$~s.}
    \tablenotetext{b}{From Winn \etal\ 2007.}
    \tablenotetext{c}{This work; both statistical and systematic uncertainties are given.}
    \tablenotetext{d}{Deviation from values in the visible.}
\end{deluxetable}

\clearpage

\clearpage

\clearpage

\clearpage

\end{document}